\documentclass[article]{aa}
\usepackage{graphicx}
\usepackage{aalongtable}
\usepackage{lscape}
\usepackage{txfonts}
\usepackage{natbib}
\bibpunct{(}{)}{;}{a}{}{,}
\usepackage{color}

\newcommand{\myarcsec}{\hbox{$.\!\!^{\prime\prime}$}}
\newcommand{\myarcmin}{\hbox{$.\!\!^{\prime}$}}
\newcommand{\myarcsecnodot}{\hbox{$\;\!\!^{\prime\prime}\;$}}

\newcommand{\angstrom}{\AA}
\newcommand{\angstromblank}{\AA$\;$}

\def\cm3{\rm ~cm^{-3}}
\def\kms{\rm ~km~s^{-1}}

\def\lsim{\!\!\!\phantom{\le}\smash{\buildrel{}\over
  {\lower2.5dd\hbox{$\buildrel{\lower2dd\hbox{$\displaystyle<$}}\over
                               \sim$}}}\,\,}
\def\gsim{\!\!\!\phantom{\ge}\smash{\buildrel{}\over
  {\lower2.5dd\hbox{$\buildrel{\lower2dd\hbox{$\displaystyle>$}}\over
                               \sim$}}}\,\,}
\begin{document}
   \title{The Crab pulsar and its pulsar-wind nebula in the optical and infrared
     \thanks{Based on observations made with the Nordic Optical Telescope, operated on the island of La Palma jointly by Denmark, Finland, Iceland, Norway, and Sweden, in the Spanish Observatorio del Roque de los Muchachos of the Instituto de Astrofisica de Canarias.)}
\thanks{Part of the data presented here have been taken using ALFOSC, which is owned by the Instituto de Astrofisica de Andalucia (IAA) and operated at the Nordic Optical Telescope under an agreement between IAA and the NBIfAFG of the Astronomical Observatory of Copenhagen.}}

   \author{A. Tziamtzis\inst{1}
          \and
           P. Lundqvist\inst{1}
          \and
           A. A. Djupvik\inst{2}
          }

   \offprints{anestis@astro.su.se}

   \institute{Stockholm Observatory, AlbaNova Science Center, Department
   of Astronomy, SE-106 91 Stockholm, Sweden
         \and
   Nordic Optical Telescope, Apartado 474, ES-38700 Santa Cruz de La Palma, Spain
}
   \date{\today }

 
  \abstract
   {}
   {We investigate the emission mechanism and evolution of pulsars that are associated with supernova remnants.}
   {We used imaging techniques in both the optical and near infrared, using images with very good seeing ($\leq$0\myarcsec6) to study the immediate surroundings of the Crab pulsar. In the case of the infrared, we took two data sets with a time window of 75 days, to check for variability in the inner part of the Crab nebula. We also measure the spectral indices of all these wisps, the nearby knot, and the interwisp medium, using our optical and infrared data. We then compared the observational results with the existing theoretical models.
}
   {We report variability in the three nearby wisps located to the northwest of the pulsar and also in a nearby anvil wisp in terms of their structure, position, and emissivity within the time window of 75 days. All the  wisps display red spectra with similar spectral indices ($\alpha_\nu = -0.58 \pm 0.08$, $\alpha_\nu = -0.63 \pm 0.07$, $\alpha_\nu = -0.53 \pm 0.08$) for the northwest triplet. The anvil wisp (anvil wisp 1) has a spectral index of $\alpha_\nu = -0.62 \pm 0.10$. Similarly, the interwisp medium regions also show red spectra similar to those of the wisps, with the spectral index being $\alpha_\nu = -0.61 \pm 0.08$, $\alpha_\nu = -0.50 \pm 0.10$, while the third interwisp region has a flatter spectrum with spectral $\alpha_\nu = -0.49 \pm 0.10$. The inner knot has a spectral index of $\alpha_\nu = -0.63 \pm 0.02$. Also, based on archival HST data and our IR data, we find that the inner knot remains stationary for a time period of 13.5 years. The projected average velocity relative to the pulsar for this period is $\lsim 8 \kms$.}
{By comparing the spectral indices of the structures in the inner Crab with the current theoretical models, we find that the Del Zanna et al. (2006) model for the synchrotron emission fits our observations, although the spectral index is at the flatter end of their modelled spectra.}
\keywords{Supernova remnant -- Crab -- pulsar -- Optical -- Infrared}

   \titlerunning{The Crab pulsar and nearby PWN in the optical and infrared}
   \authorrunning{A. Tziamtzis et al.}

   \maketitle


\section{Introduction}

The Crab nebula is one of the most studied targets in the sky. It is a remnant from a supernova explosion that occurred in 1054. The Crab pulsar (m$_V$$\sim$16), the first pulsar with optically detected pulses (Cocke et al. 1969), is located at the centre of the nebula, and is responsible for powering the nebula. The pulsar, together with the filaments and the continuum emitting pulsar-wind nebula (PWN), form the Crab nebula. The Crab nebula was the first detected astrophysical source of synchrotron radiation, as suggested by Shklovsky (1953) and was confirmed by polarisation observations a year later by Dombrovsky (1954). The inner part of the Crab nebula is a region consisting of jets, a torus of X-ray emission (Aschenbach \& Brinkmann 1975), and complexes of sharp wisps (Scargle 1969). The structures give a unique opportunity to study the pulsar, along with its wind and the interaction it has with the rest of the remnant. 

Evidence of activity in the Crab pulsar and its surroundings has existed for a few decades. Variability in the structures around the pulsar were mentioned by Lampland (1921) and by Oort \& Walraven (1956). A very comprehensive study has been done by Scargle (1969), who found that the wisps may show relativistic motion, but conversely always seem to be more or less located at the same position. Using the WFPC2 onboard HST high resolution surveys were conducted by Hester et al. (1995; 2002). These data were complemented with X-ray data from ROSAT and Chandra to get a complete picture of the Crab pulsar and its immediate surroundings. The most outstanding discovery is the presence of two knots that are situated 0\myarcsec65 and 3\myarcsec8 southeast of the pulsar. The X-ray data have also shown many knots around the pulsar that show variability with time. For a complete look at the Crab pulsar and its nebula see Hester (2008).

Sollerman (2003) used ISAAC on the VLT, to conduct an IR survey on JHK${\rm_s}$ bands in order to investigate the Crab pulsar wind nebula further. Using the VLT data, together with archive HST optical data he measured a spectral index for the knot equal to $\alpha_{\nu} = -0.8$. Melatos et al. (2005) went further into investigating the Crab PWN using IR images to check for short time-scale variations around the Crab pulsar. They found variations in the emission of the Crab wisps on a time scale of 1.2 ksec of $\pm$24\%  $\pm$ 4\% in the K band and $\pm$14 $\pm$5\% in the J band. Apart from the variation estimations, Melatos et al. (2005) calculated the spectral indices from the structures near the pulsar, finding values for the spectral index that vary from $\alpha_{\nu} = -0.16$ for the rod (this the brightest feature in the shock located $\sim$ 4\myarcsecnodot in a southeastern orientation from the pulsar) down to $\alpha_{\nu} = - 0.76$ for the innermost knot.

Despite the amount of work that has been done on the Crab, there are still many unanswered questions regarding its nature and, in general, about the supernova explosion mechanism and how the pulsar emission mechanism operates and evolves. Also there is not any model that can fully explain the observed features of the emitted radiation, especially in the inner part of the PWN. In this article we present our near infrared and optical survey that we conducted using the Nordic Optical Telescope in La Palma, Spain, with a gap of two and a half months (in the case of the NIR) in order to check for variability in the enviroment around the Crab pulsar. In Sect. 2 we present our observing procedure followed by the data reduction process. In Sect. 3 we present our photometric results, and in Sect. 4 we present arguments for the nature of the wisps and the knots. We make our conclusions in Sect. 5.

\begin{figure}[ht]
\includegraphics[width=1.0\hsize]{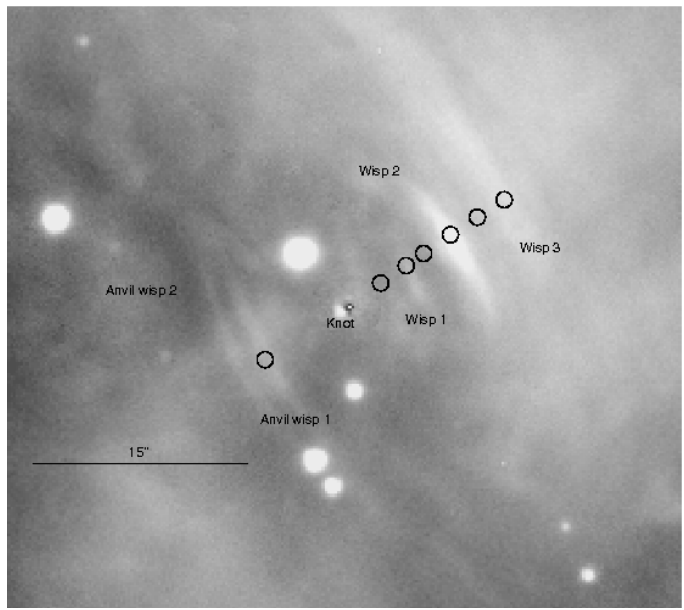}
\caption{I-band image of the Crab nebula taken with ALFOSC at the NOT on 7 December 2007. The pulsar has been removed with the PSF to reveal the nearby knot that lies at a distance of 0\myarcsec65. The field of view is $\sim$ 45$\times$45 arcseconds, with north pointing upwards and east to the left. The areas that we used to measure the emission from the wisps and the interwisp regions are marked with circles.}
\end{figure}

\section{Observations}  

\begin{figure}[t]
\includegraphics[width=0.33\hsize]{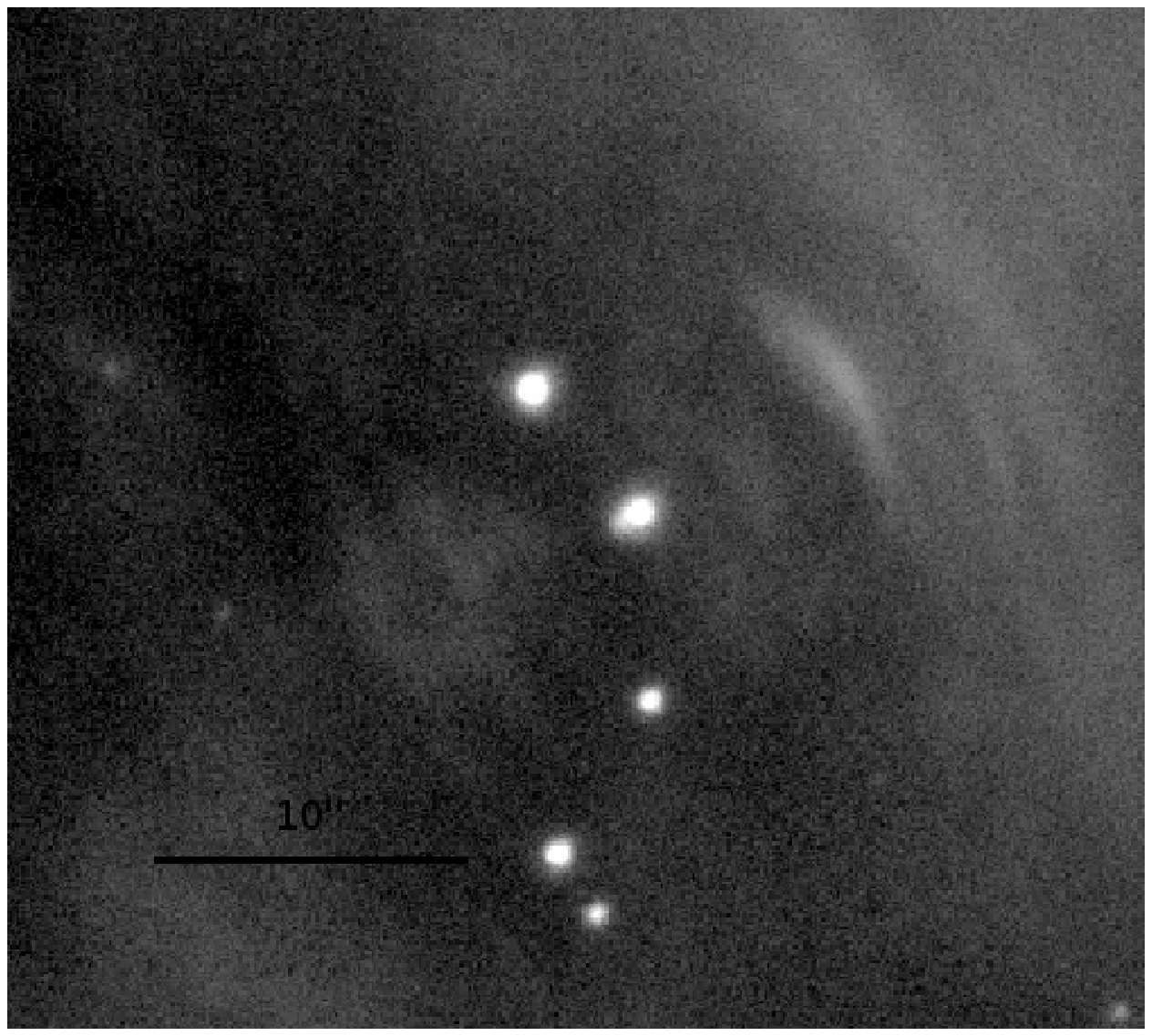}\hspace*{1mm}\includegraphics[width=0.33\hsize]{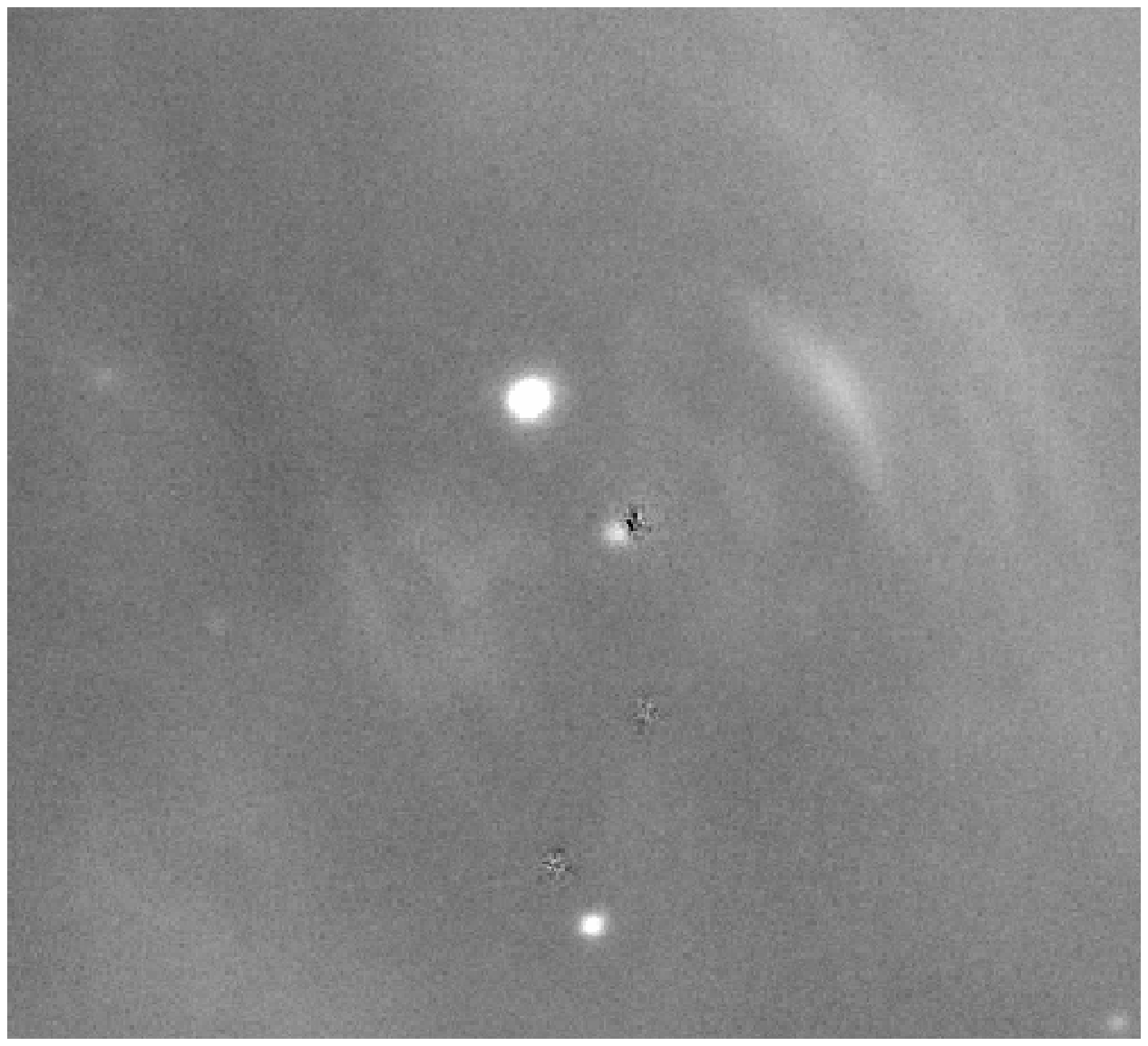}\hspace*{1mm}\includegraphics[width=0.34\hsize]{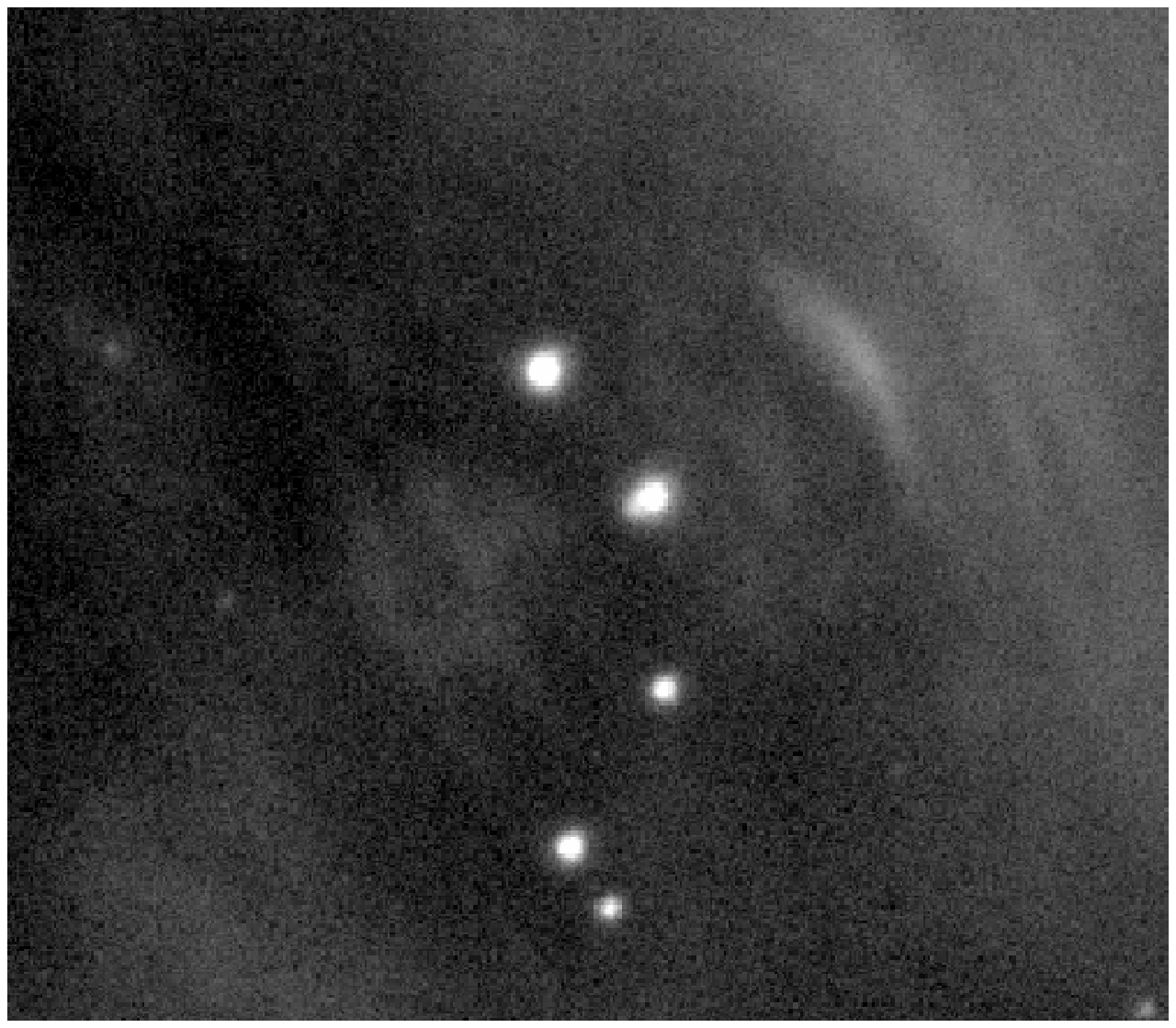}
\includegraphics[width=0.33\hsize]{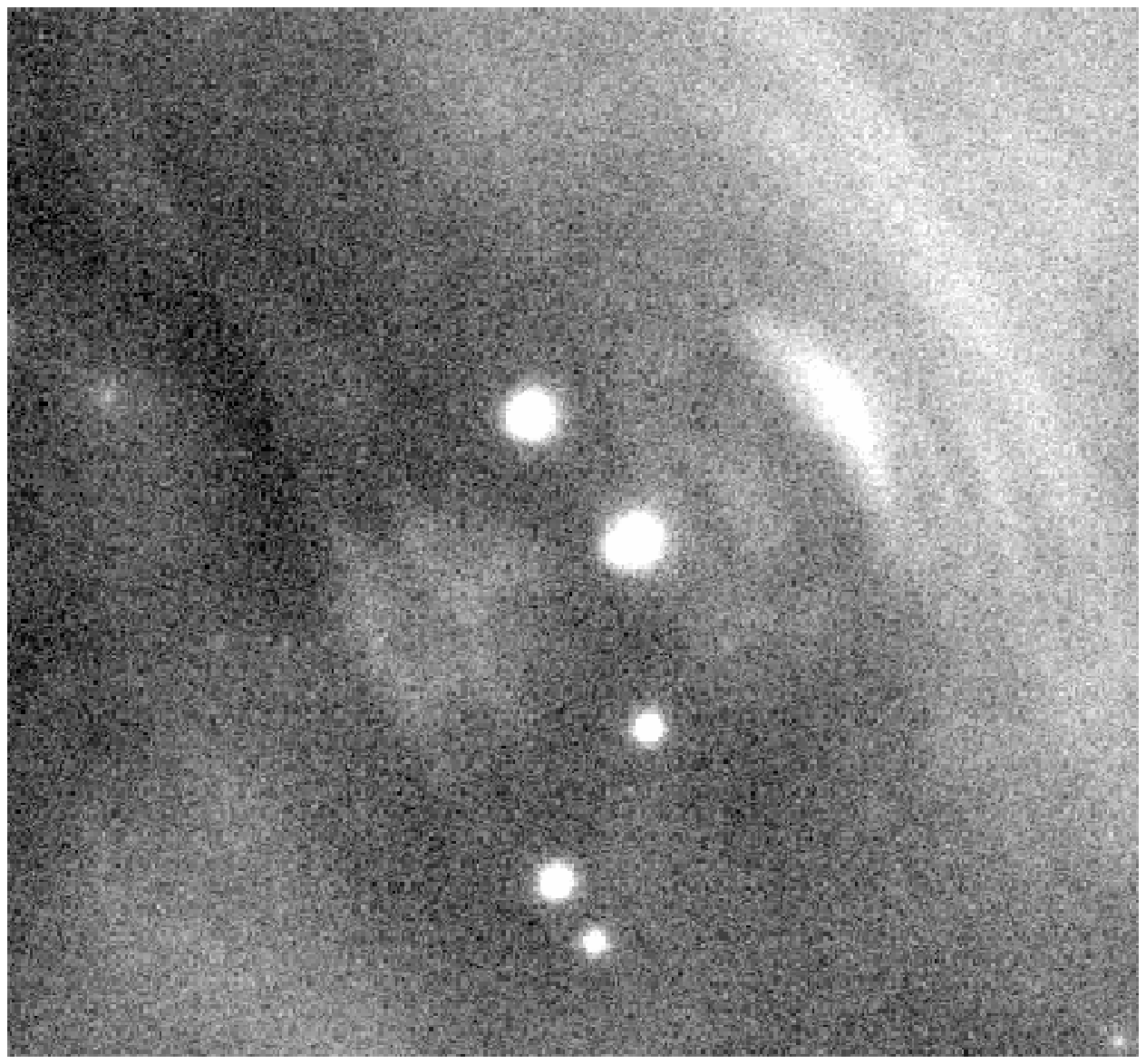}\hspace*{1mm}\includegraphics[width=0.33\hsize]{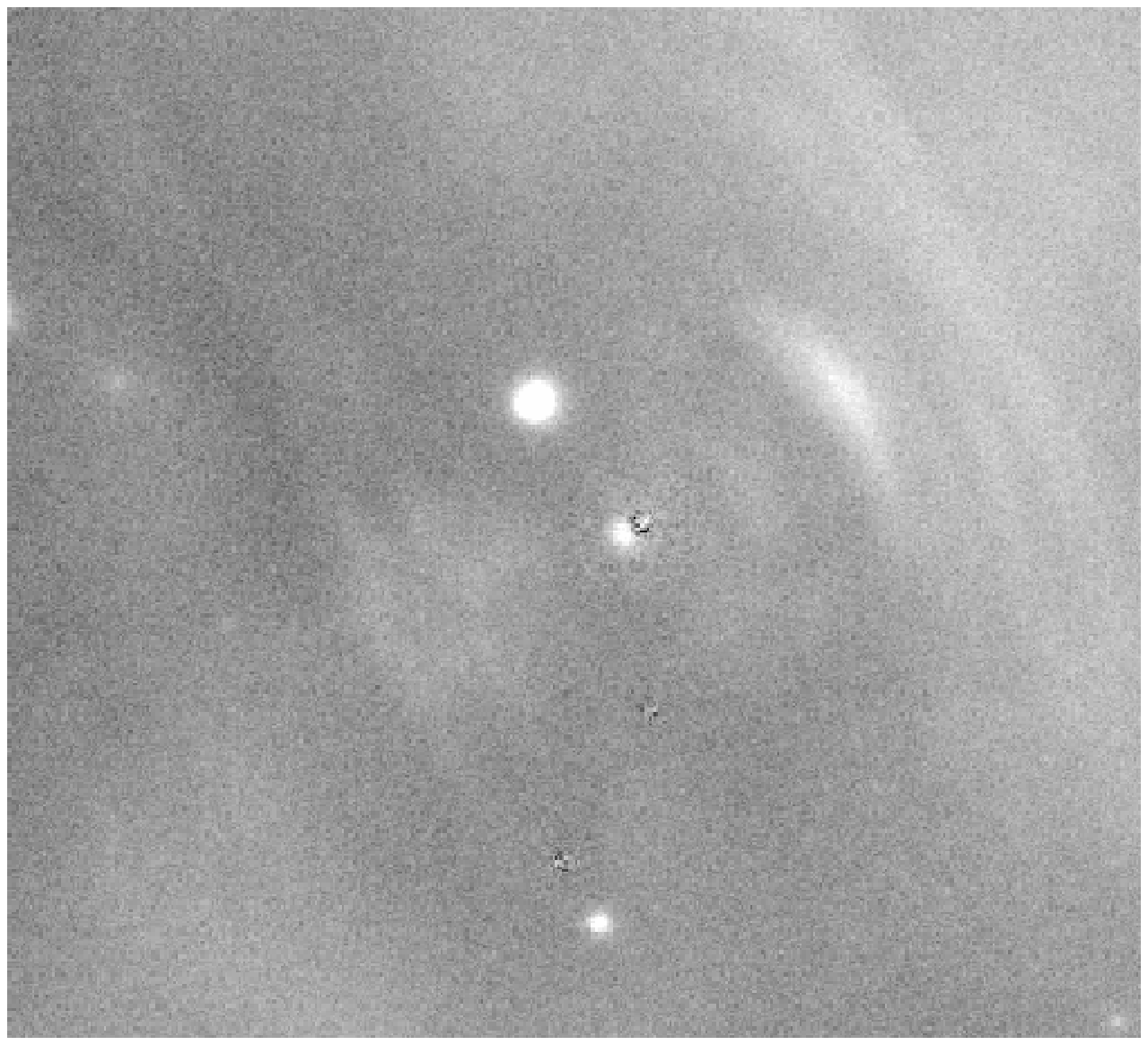}\hspace*{1mm}\includegraphics[width=0.33\hsize]{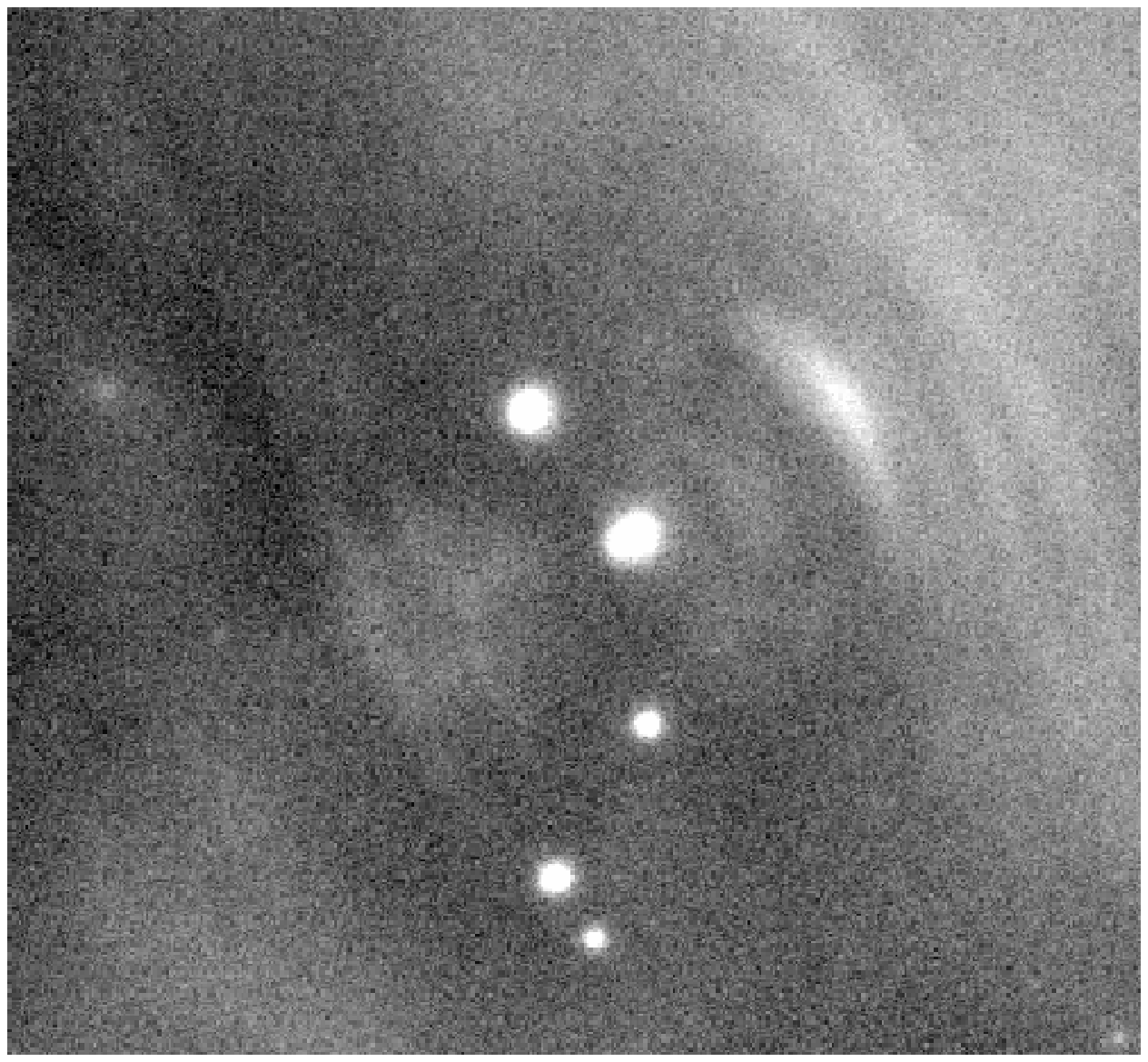}

\caption{{\it Upper panel:} H-band images of the inner part of the Crab nebula taken with the high-resolution camera of NOTCam (0\myarcsec078/pix) at the NOT on 30 September and 13 December 2007. The pulsar appears to be elongated due to the presence of the nearby (0\myarcsec65) knot. The wisps that surround the pulsar are prominent on both frames. By setting the cuts in an appropriate manner, the knot can also be visible in these images. The knot can be clearly seen after the subtraction of the pulsar on the image from 30 September (middle). On the right side of the panel, we show the H band image from the night 13 December, where the changes in the wisp triplet are obvious in terms of brightness and structure especially for the outer wisp. Similar changes can be seen in the anvil region too. (See Fig. 1 for a description of the Crab nomenclature)
{\it Lower panel:}  K${\rm_s}$ band images of the inner part of the Crab nebula from the same nights. On the left side of the panel, we show the image from the night of 30 September. In the middle of the panel we show in the inner knot after we subtracted the pulsar with the PSF, and finally in the right side the image from the night of 13 December. In all images, the field of view is $\sim$ 35$\times$32 arcseconds, and north is up, east is to the left.}
\end{figure}
 
The NIR observations were carried out on two seperate occasions. The first took place on 30 September 2007 and the second on 13 December 2007. Both runs were done using the high-resolution camera of NOTCam\footnote[1]{http://www.not.iac.es/instruments/notcam/} on the Nordic Optical Telescope (NOT) in Observatorio del Roche de los Muchachos in La Palma Spain. The observations used an H-band filter ($\lambda_c$ = 1.626 $\mu$m, FWHM = 0.296 $\mu$m) and a filter in the K${\rm_s}$ band ($\lambda_c$ = 2.140 $\mu$m FWHM = 0.310 $\mu$m). The detector of the NOTCam is an 18.5 $\mu$m$\times$ 1024 $\times$ 1024 HgCdTe (Hawaii) array. The pixel scale of the HR camera is 0\myarcsec078/pix which gives a field of view of ~ 80\myarcsecnodot $\times$ 80\myarcsecnodot.  The first observing run used the Engineering Grade Array, while the second run used the new Science Grade Array for NOTCam.

The observations were performed in the same manner for both NOTCam runs. Because the whole field of view of the HR camera contains extended emission from the nebula, we used beam-switching between target and sky, located 180\myarcsecnodot  towards the south, in order to account for a proper background subtraction. Each position (ON target, as well as OFF target) was exposed for 80 seconds using a ramp-sampling readout mode where the array is read nondestructively every 10 seconds. A linear regression analysis is performed on the 8 readouts to give the final image.

We achieved a total exposure time of 1500 sec for the H band and 750 sec in the K${\rm_s}$ band, partly limited by weather on the first night. The average seeing on the night of 30 September was 0\myarcsec45 and the data were taken under photometric conditions. On the second night, due to variable seeing (0\myarcsec50 $\leq$ seeing $\leq$ 0\myarcsec80), we selected the sharpest images only ($\leq$ 0\myarcsec65), giving an effective total exposure time of 640 sec in the H band and 800 sec in the K${\rm_s}$ band. The seeing in the coadded IR images on the night of 30 September is $\sim$ 0\myarcsec45, while for the night of 13 December is $\sim$ 0\myarcsec55. On both nights we observed the photometric standard star AS05 (Hunt et al. 1998).

Optical observations were carried out on 7 December 2007 (I band) and 20 December 2007 (U \& V) using the Andalucia Faint Object Spectrograph and Camera (ALFOSC) at NOT\footnote[2]{http://www.not.iac.es/instruments/alfosc}. ALFOSC has a pixel scale of 0\myarcsec19 per pixel and it offers a field of view of 6\myarcmin5$\times$6\myarcmin5. A series of frames were taken using an I ($\lambda_c$ = 7970 \angstromblank FWHM = 1570 \AA) continuum filter in the night of 7 December. In total we exposed for 2900 sec in the I band in a 13$\times$200sec procedure plus a 300 sec exposure. The U band images on 20 December 2007 were taken with a broad U filter ($\lambda_c$ = 3550 \angstrom, FWHM = 550 \angstrom) with a total exposure time of 3600 sec. For the images in the V band we used a Stromgren filter ($\lambda_c$ = 5470 \angstrom, FWHM = 210 \angstrom) with a total exposure time of 2700 sec. We chose these filters since, according to their profiles, there is no contamination from any strong emission line (e.g., H$\alpha$ or O [III]). To reveal the nearby (i.e., $\sim$ 0\myarcsec6) knot we used only the exposures with  seeing below 0\myarcsec6. Due to seeing limitation the knot can not be revealed in the U and V bands images, so we used them just to study the wisps. For the purposes of flux calibration of our data, we used a spectrum of the pulsar that we took from Sollerman et al. (2000).

Images in the V band ($\lambda_c$ = 5480 \AA, FWHM = 485 \AA) taken with the WFPC2 with total exposure time of 2000 sec, obtained on 14 August 1995 were requested from HST science archives to check that the knot next to the Crab pulsar has moved during the 13-year period.

\begin{table}
\caption{Observational log.}
\label{expdates}
\centering
\begin{tabular}{c c c c c l}
\hline\hline
Source & Filter & Exp. time & Eff. Exp. Time & Seeing & Date\\
       &      & (Seconds) & (Seconds) & (Arcsec)  &        \\
\hline
\hline
M1 & H & 1500 & 1500 & 0\myarcsec45 & 070930\\
M1 & K${\rm_s}$ & 750 & 750 & 0\myarcsec45 & 070930\\ 
M1 & H & 1500 & 640 & 0\myarcsec55 & 071213\\
M1 & K${\rm_s}$ & 1500 & 800 & 0\myarcsec55 & 071213\\
M1 & V & 2000 & 2000 & 0\myarcsec10 & 951406\\
M1 & I & 2900 & 400 & 0\myarcsec55 & 071207\\
M1 & U & 3600 & 3600 & 1\myarcsec20 & 071220\\
M1 & V & 2700 & 2700 & 0\myarcsec80 & 071220\\
AS05 & H & 250 & 250 & 0\myarcsec40 &  070930\\
AS05 & K${\rm_s}$ & 250 & 250 & 0\myarcsec30 & 070930\\
AS05 & H & 250 & 250 & 0\myarcsec70 &  071213\\
AS05 & K${\rm_s}$ & 250 & 250 & 0\myarcsec70 & 071213\\
\hline
\end{tabular}
\end{table}

\subsection{Data reduction}

The near-IR images were reduced using IRAF\footnote{http://iraf.noao.edu/docs/photom.html} and a set of our own scripts. All the images were flat-fielded by using differential skyflats. These can be created by taking two sets of flat fields, one set with high counts (10,000-15000 ADU) and another with low counts (700-1500 ADU) if these have been obtained within a limited time span, which is the case for the right window during twilight. These frames are used to create master flats, and the differential approach ensures that the thermal contribution is subtracted. Another benefit of obtaining these series of frames is that they can be used to determine the bad pixels in the array. For cosmetic reasons we applied a bad pixel mask to remove bad and warm pixels from our science frames. Sky subtraction was done by subtracting the master sky frame that we made from the individual sky exposures we took. The master sky was properly scaled before subtracting it from each target image. Then all the dithered target images were flat-field corrected, aligned, shifted, and median combined.

The photometric zeropoints of the two infrared filters were obtained from the photometric standard AS05 (Hunt et al. 1998), and they are found to be 23.70 $\pm$0.01 mag. for the H band and 23.16 $\pm$0.01 mag. for the K$_s$ band for the data that we got on 30 of September and 24.10 $\pm$0.01 mag. for the H band and 23.58 $\pm$0.01 mag for the K$_s$ for the data from 13 December. (The difference in the photometric zero points is due to the fact that we used two different arrays. Also the estimated zeropoints are for 1 electron per second) The magnitudes of the pulsar on both occasions  were calculated with the standard PSF subtraction method within IRAF's DAOPHOT (Stetson 1987). Our H and K${\rm_s}$ magnitudes of the pulsar give  slightly lower fluxes than the 2MASS catalogue by 0.09 and 0.14 magnitudes in the H and K$_s$ band, respectively (see Table 2) for 30 of September and 0.05 and 0.08 magnitudes for 13 of December. We believe that our PSF fitting has good accuracy at both epochs since the knot east of the pulsar is clearly visible at both epochs. The errors that we measured for the pulsar, as given by IRAF's ALLSTAR, are lower than 0.05 magnitudes for both filters. This also  includes the 0.02 magnitudes uncertainty estimated from the aperture correction. The ALLSTAR package is the most suitable package for varying enviroments like the one of the Crab pulsar so no other methods for the PSF subtraction were tested.

For the reductions of optical data we used THELI,\footnote{http://www.astro.uni-bonn.de/~mischa/theli.html} which is an automatic pipeline for the reduction of optical and infrared data (Erben et al. 2005). The data were bias-subtracted and flat-field corrected. Because of the dithering that we applied, we also made weight maps for each frame in order to monitor each pixel separately. We then applied an astrometric solution to our frames and removed the background on each frame by using the images themselves by subtracting a constant value. We then shifted and coadded the images. 

Once again, the complex enviroment around the Crab pulsar required  constructing PSF models to measure the magnitudes of the pulsar in the U, V, and I bands and to isolate the knot in the I band.  For the flux calibration of the data, we used a spectrum of the pulsar from Sollerman et al. (2000). For the I-band images from 7 December, we found a photometric zeropoint of 25.35 mag from which we estimated the magnitude of the pulsar to be 16.15 mag. For the U and V bands, we estimated  zeropoints of 24.90 mag and 23.75 mag, respectively, and we found the magnitudes of the pulsar to be 17.55 mag and 16.80 mag, respectively. In all three filters (U, V, and I) the estimated uncertainty is around 0.02 mag. We also applied the needed corrections for the effect of atmosperic extinction.

\begin{table}
\caption{Magnitudes of the Crab pulsar.}
\label{expdates}
\centering
\begin{tabular}{l c c c}
\hline\hline
Source & Date & H-band & K${\rm_s}$-band \\
\hline
\hline
M1 & 2007/09/30 & $14.14 \pm 0.02$ & $13.64 \pm 0.05$ \\
M1 & 2007/12/13 & $14.10 \pm 0.02$ & $13.58 \pm 0.05$ \\ 
\hline
\end{tabular}
\end{table}

\begin{figure}[t]
\includegraphics[width=1.0\hsize]{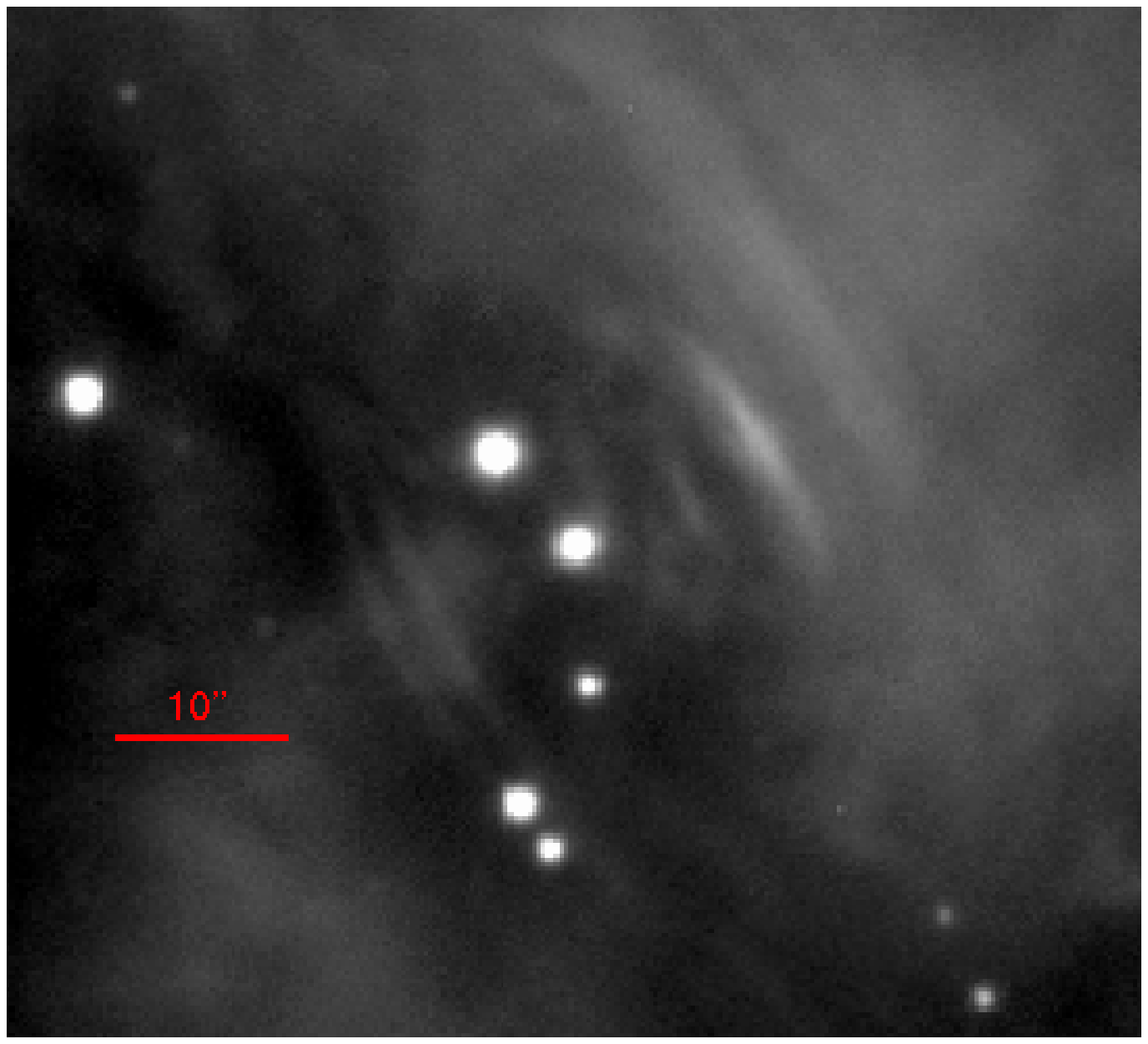}
\includegraphics[width=1.0\hsize]{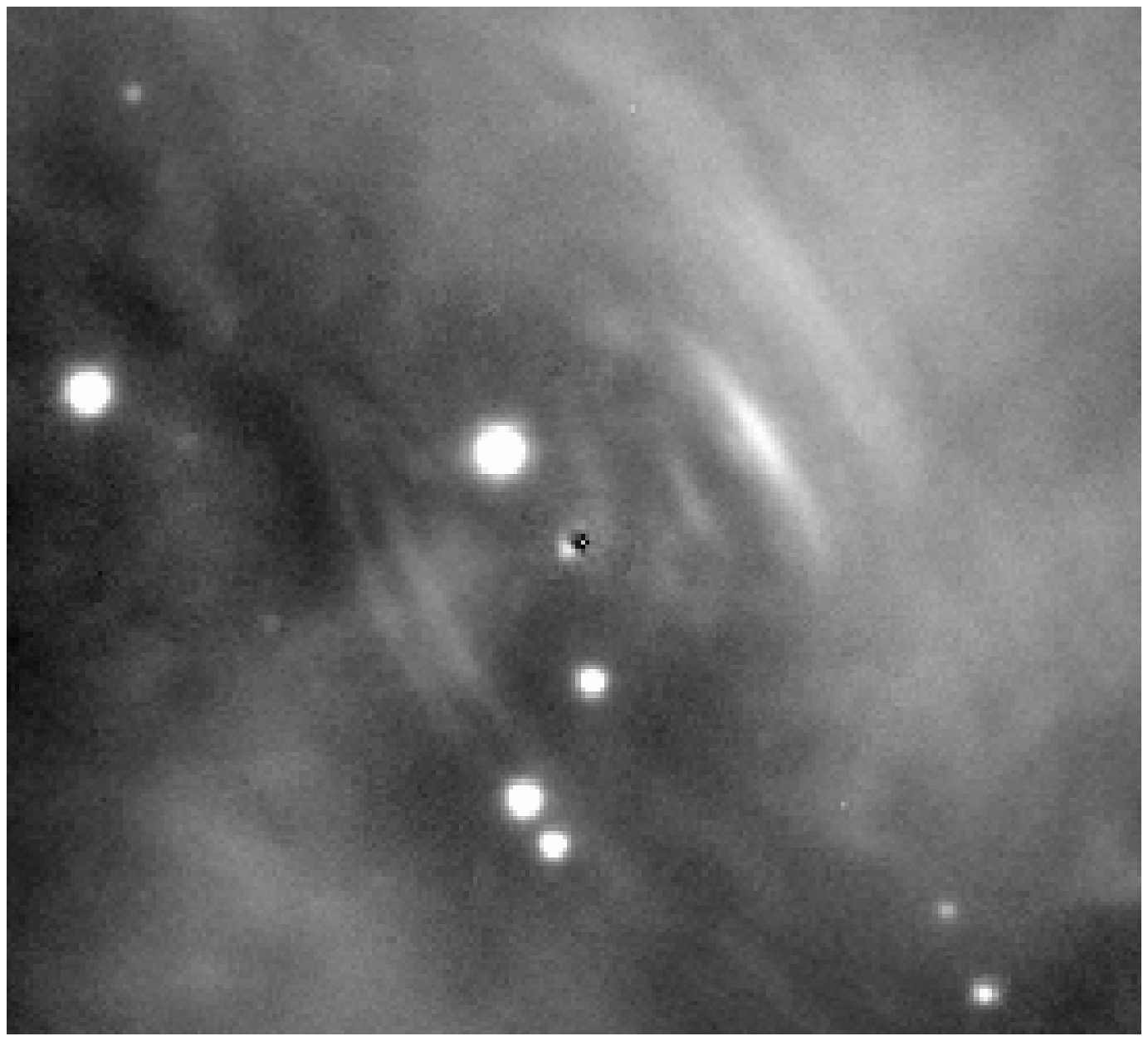}
\caption{{\it Top:} I band image of the Crab pulsar taken with NOT/ALFOSC on the night of 7 December. {\it Bottom:} The same image after the subtraction of the pulsar with the PSF. The field of view is approximately 45$\times$45 arcseconds. The residual emission southeast of the pulsar's position is the nearby knot first reported by Hester et al. (1995). On both images, north is up and east to the left.}
\end{figure}

\section{Analysis of photometry}

\subsection{The inner nebula}

 Our images show most of the known features that are present in the inner part of the Crab nebula. In all our frames we see the wisps in great detail. In the case of the H band image, there is some contamination by the [Fe~II] 1.64 $\mu$m line detected by Graham et al. (1990). The K${\rm_s}$ band image is dominated by the synchrotron emission, but a small amount of the emission comes from the molecular hydrogen 1$\rightarrow$0 S(1) 2.122 $\mu$m and the Br$\gamma$ 2.163 $\mu$m lines. (Use Fig. 1 to get an orientation of the photometric analysis that we made in the inner part of the Crab nebula).

\subsection{The pulsar and the knot.}

The pulsar appears to be elongated on all images with good seeing ($\leq$ 0\myarcsec6). This due to the presence of the nearby knot that was first identified by Hester et al. (1995) on the HST/WFPC2 images. Hester et al. (1995) mentions the presence of a second knot near the pulsar, in the same orientation as the first, but lying closer to the anvil at a distance of 3\myarcsec8 from the pulsar. From our data set there is no evidence of this knot since it is dimmer than the first one by a factor of almost five (Hester et al. 1995). 

After subtracting the pulsar with the PSF we fitted an aperture in order to measure the flux of the nearby knot. In the case of the K${\rm_s}$ band the flux of the knot is $\sim 10$\% of the flux of the pulsar. This shows that in case of bad seeing conditons one would measure a brighter pulsar. For the H band the flux of the knot is $\sim 8$\% of the flux that we measured for the pulsar. The measured fluxes are 2.80 $\pm$ 0.15 $\times10^{-27}$ ergs sec$^{-1}$cm$^{-2}$Hz$^{-1}$ for the K${\rm_s}$ filter and 2.65 $\pm 0.15$ $\times10^{-27}$ ergs sec$^{-1}$cm$^{-2}$Hz$^{-1}$ for the H filter. Two and a half months later we found that the emitted flux from the knot is constant. The measurements from the night of 13 December gave fluxes of 2.90 $\pm$ 0.10 $\times10^{-27}$ ergs sec$^{-1}$cm$^{-2}$Hz$^{-1}$ for the K${\rm_s}$ filter and 2.75 $\pm$ 0.10 $\times10^{-27}$ ergs sec$^{-1}$cm$^{-2}$Hz$^{-1}$. Below we discuss our observations and results.

The knot is not circular in shape, but has an arc-like shape. The latter can be used as an argument that it is not a background or foreground star and that it is indeed a structure associated with the pulsar. Another feature that supports this scenario is that this knot, together with the second knot that Hester et al. (1995) mentioned, is aligned with the X-ray jet emanating from the pulsar. In the I band images, this is more pronounced since we see the knot aligned with an outflow of gas (optical jet) that goes through the anvil (see Fig. 2). Also its morphology indicates that it has to be a shock wave or some instability feature in the flow that is responsible for the jet. 
  
\begin{table}
\caption{Proper motion of the Crab pulsar.}
\label{expdates}
\centering
\begin{tabular}{l c c c}
\hline\hline
Date & Knot & Star 1 & Star 2 \\
\hline
\hline
94/08/14 & 0\myarcsec625(0.005)\footnote{3$\sigma$ pointing uncertainty}  & 5\myarcsec00 & 4\myarcsec94 \\
07/09/30 & 0\myarcsec635(0.012)  & 5\myarcsec22 & 5\myarcsec13 \\ 
\hline
\end{tabular}
\end{table}

In the I band images we only managed to get two satisfactory frames in terms of seeing during the night of 7 December to extract the knot near the pulsar (see Fig. 3). Once again we managed to reveal the knot after the removal of the pulsar with the PSF. The structure of the knot appears to be the same as in the H and K${\rm_s}$ images. The measured distance is again found to be $\sim$ 0\myarcsecnodot65, and the flux that we measured from the knot in the I band is 1.60 $\pm$ 0.10 $\times$10$^{-27}$ ergs sec$^{-1}$cm$^{-2}$Hz$^{-1}$, while the pulsar has a flux of 2.45$\times$10$^{-26}$ ergs sec$^{-1}$cm$^{-2}$Hz$^{-1}$. The flux of the knot therefore corresponds to $\sim 6.5$\% of the flux of the pulsar.

We used HST/WFPC2 archival images to study a possible motion of the red knot with regard to the background/foreground stars and the pulsar. Data that date back 13.5 years from our observing run were used for this purpose. According to the current measurements the Crab pulsar has proper motion of 15 mas yr$^{-1}$ (Kaplan et al. 2008). This means that the pulsar in this amount of time has moved $\sim$ $0\farcs2$. We measured the distances between the pulsar and the knot at both epochs, and the two results agree (cf. Table 3). If the knot was a background/foreground source, then its proper motion would have been different than the one of the pulsar. This is also supported by the comparison that we made with stars close to the pulsar. In Table 3, we show the relative distance between the pulsar and the knot, as well as two nearby stars. The pulsar-knot distance has not changed, whereas the Crab has moved about $0\farcs2$ relative to the stars, as expected. The distance between the knot and the pulsar that we measure is very close to the values estimated $\sim 0\farcs65$) by Hester et al. (1995) and Sollerman (2003). 

From our measurements we can estimate an upper limit for any possible relative motion in the pulsar-knot system. By using the relation $V_{pm} = 4.74 \mu D$ where V is transverse velocity in units of kms$^{-1}$ (Lyne \& Lorimer 1995), one can estimate the shifting of the pulsar, where $\mu$ is the proper motion per year in mas and $D$ is the distance in kpc. Using a distance of 2 kpc (Kargaltsev \& Pavlov 2008) for the Crab and a proper motion of 15 mas/yr (Kaplan et al. 2008), the projected velocity of the pulsar is $\sim 142 \kms$. The motion appears to be aligned with the jet axis, the pulsar moving in a northwest direction towards the wisp triplet. By using the uncertainties from the separation estimation between the pulsar and the knot, we estimate that the upper limit to their relative projected velocity is not more than $\sim 8 \kms$, which therefore makes it safe to state that the knot is quasistationary with regard to the pulsar, at least over a period of 13.5 years. Thus, assuming that the knot lies along the jet axis and that the jet axis is tilted $\sim 30$ degrees to the plane of the sky (Hester 2008), as well as taking into account that the pulsar also moves along the jet axis, the knot trails the pulsar by $\lsim 10 \kms$, corresponding to $\lsim 6\%$ of the space motion of the pulsar.

\begin{table*}
\caption{Infrared results from the Crab PWN.}
\label{catalog}
\centering
\begin{tabular}{c c c c c c c}
\hline \hline
Region & Filter &  Flux(September 30) & Flux(December 13)  & Distance(September 30)\footnote{In arcseconds from the pulsar} & Distance(December 13)\\
\hline
Knot   & H & 2.65(0.15)\footnote{Flux in units of ergs/s/cm$^2$/Hz$\times10^{-27}$.} & 2.75(0.10)  & 0.65 & 0.65\\
Wisp 1 & H & 3.15(0.20)\footnote{1$\sigma$ uncertainty in the same units.} & 3.95(0.20)  & 4.20 & 6.00\\
Wisp 2 & H & 5.30(0.30)\footnote{We used $E(B-V) = 0.52$ and $R_V = 3.1$ for derreddening.} & 6.25(0.25)  & 8.40 & 9.70\\
Wisp 3 & H & 4.20(0.25) & 4.75(0.20)  & 13.60 & 14.00\\
Anvil Wisp 1 & H & 3.20(0.15) & 3.60(0.15)  & 5.30 & 6.30\\
Knot   & K${\rm_s}$ & 2.80(0.15) & 2.90(0.10)  & 0.65 & 0.65\\
Wisp 1 & K${\rm_s}$ & 3.50(0.25) & 4.45(0.20)  & 4.20 & 6.00\\
Wisp 2 & K${\rm_s}$ & 5.55(0.30) & 6.45(0.25)  & 8.40 & 9.70\\
Wisp 3 & K${\rm_s}$ & 4.50(0.25) & 4.90(0.20)  &13.60 & 14.00\\
Anvil Wisp 1 & K${\rm_s}$ & 3.35(0.20) & 4.10(0.15) &5.30 & 6.30\\
\hline
\end{tabular}
\end{table*}

\subsection{Time variable changes in the inner nebula}

By looking at our first data set we see that there is a change in the region where the anvil is present (See Fig. 2 for a complete view). An obvious difference from the HST/WFPC2 V band image from Hester et al. (1995) is that we see two clear wisps in the anvil region, and we also see a presumed outflow of gas that is aligned with the X-ray jet (see also Fig. 3). The northwest direction of the pulsar shows the known features. We see the three wisps that are separated from each other by just a few arcseconds. The nearby wisp is dim on the images taken on 30 September. Wisp 2 is the brightest one followed by the outer wisp that shows some structure, where it is obvious that its edges are breaking. Two and a half months later, it seems that the enviroment around the pulsar had changed significantly. We cleary see that the nearby wisp west of the pulsar appears to have brightened and also shifted outwards. Changes are also seen in the wisps that are further away in the western direction. The brightest wisp (wisp 2) appeared to have a small change in its thickness, and its brightness had increased by a factor of $\sim$1.20 in the H band, while the one that lies the furthest away, seems to have a single structure now. On average the flux in the H band is higher by a factor of $\sim$ 1.20 on the night of 13 December. The flux on the K${\rm_s}$ band increased also by the same factor. The two wisps in the anvil region are more clearly seen now on both bands, and in the case of the H band image we see that the pulsar is surrounded by a halo, which appears to be almost 30\myarcsecnodot across. Other evident features in the images of 13 December are the outflow of gas that goes through the anvil region and a low surface brightness gas region just next to the pulsar (orientated to the west). In Table 4 we show all the photometric measurements from the wisps on both epochs and also the distance that the wisps have from the pulsar (See Fig. 4 to see the shift of the wisps between the two epochs). For the galactic extinction, we assumed $R_V = 3.1$ and $E(B-V) = 0.52$ using Cardelli et al. (1989). 

\begin{figure}[t]
\includegraphics[width=1.0\hsize]{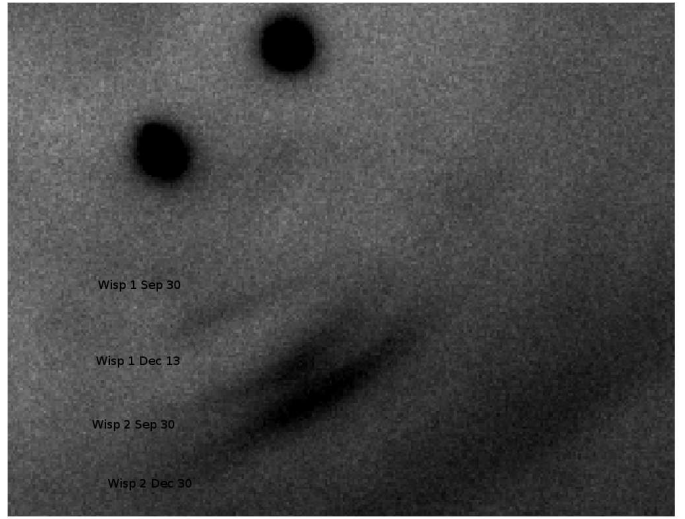}
\caption{Stacked H band image from our two observing runs on 30 September and 13 December. This shift corresponds to 1\myarcsec2 or 2800AU. Such a movement corresponds to a speed that is $\sim 0.2c$. North is to the right and west is up.}
\end{figure}

\subsection{Spectral indices from the inner enviroment of the Crab}

From the ALFOSC and NOTCam (13 December) data we measured the spectral index of the knot, the anvil (wisp 1), and the northwest wisp triplet, and we compared our results with the spectral index of the Crab pulsar and with the regions between the wisps (interwisp medium). It is well known that the Crab pulsar has a flat spectrum. The pulsar emission from $\gamma$-rays to IR is produced by synchrotron radiation (Lyne \& Graham 1998, and references within). Electrons following a power law N(E)dE = CE$^{-\gamma}$dE, where E is the energy, C a constant, and $\gamma$ the electron spectral index, will produce synchrotron radiation that obeys a power law $F_\nu = K(\nu/\nu_o)^{\alpha_\nu}$ ergs s$^{-1}$ cm$^{-2}$ Hz$^{-1}$. For synchrotron radiation, the photon spectral index $\alpha_\nu$ is related to the electron spectral index $\gamma$ via  $\alpha_\nu = -(\gamma - 1)/2$.

 For our measurements, we chose the same regions in all the images for all three wisps and we measured the surface brightness there. These regions are located on the jet axis. (See Fig. 1 for orientation, and Tables 5 and 6 for results.) A spectrum of the pulsar taken from Sollerman et al. (2000) and our photometric data were dereddened using $R_V = 3.1$ and $E(B-V) = 0.52$. By performing a linear fit to both data sets, we found that the spectral index of the pulsar is  $\alpha_\nu =  0.12$, while the wisps have spectral indices equal to $\alpha_\nu = -0.58\pm0.08$ for wisp 1, $\alpha_\nu = -0.63\pm0.07$ for wisp 2 (see Fig. 5), and wisp 3 has a spectral index of  $\alpha_\nu = -0.53\pm0.08$. Thus we see that the wisps have a rather red spectrum that is similar to the spectrum of the knot for which we found a spectral index of  $\alpha_\nu = -0.63\pm0.02$. Despite using only three points for determining the spectral index of the knot, this resemblance can be used to give additional support to the pulsar-knot association, since the three wisps and the knot show a spectrum with similar behaviour. Finally, the spectral index from the anvil region (anvil wisp 1) also shows a red spectrum similar to those of the other wisps. We measured a spectral of $\alpha_\nu = -0.62\pm0.10$. 
   
We also measured the flux in the regions between the wisps in our frames in order to compare the values with the measured wisp fluxes and find that these regions have a spectrum similar to that of the wisps and the knot ($\alpha_\nu = -0.61\pm0.08$, $\alpha_\nu = -0.50\pm0.08$, $\alpha_\nu = -0.49\pm0.07$), although the surface brightness is of course lower than for the wisps. 

Veron-Cetty \& Woltjer (1993), who made a spectral index map of the whole Crab nebula in the optical by dividing it into small 10\myarcsecnodot $\times$ 10\myarcsecnodot boxes (see their Fig. 1), found a variation in the spectral index within the PWN. Their results showed an increase of $\alpha_{\nu}$ as the distance increases from the pulsar, starting from values of $\alpha_{\nu} = - 0.57$ near the pulsar and decreasing to $\alpha_{\nu} = -1$ at the edge of the synchrotron nebula. Veron-Cetty \& Woltjer (1993) report values for $\alpha_{\nu}$ for the region of the Crab PWN that we focussed on in the range $-0.60$ and $-0.57$, which is in good agreement with our results and which indicates that the spectrum has not changed since their measurements.

\begin{figure}[t] 
\includegraphics[width=1.0\hsize]{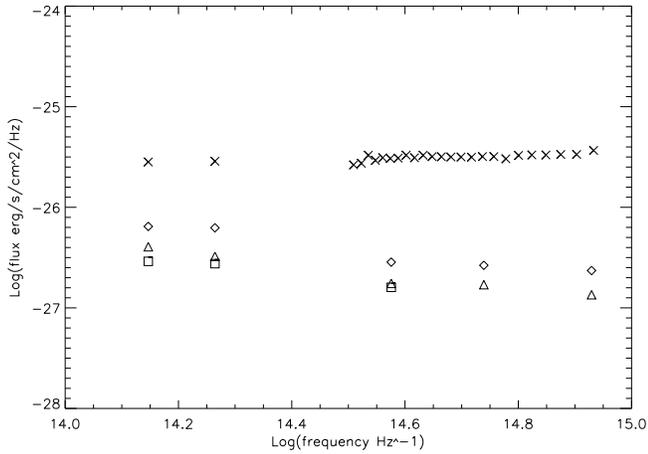}
\caption{Spectra of wisp 2, the inner knot, together with spectra from the pulsar and the synchrotron nebula (Interwisp 1) for comparison. The pulsar spectrum (represented by X) is taken from Sollerman et al. (2002). The two infrared points in the spectrum of the pulsar are from our infrared observations from 13 December. The spectrum of the knot is shown by squares, the spectrum of wisp 2 is represented by diamonds, and the synchrotron nebula (Interwisp 1) is represented by triangles. All the data have been corrected for extinction using $R_V = 3.1$ and $E(B-V) = 0.52$.}
\end{figure}

\begin{table*}
\caption{Photometric results from the Crab Wisps and the knot}
\label{catalog}
\centering
\begin{tabular}{c c c c c c c c c}
\hline \hline
Filter & Inter Wisp 1 & Wisp 1 &  Inter Wisp 2 & Wisp 2 & Inter Wisp 3 & Wisp 3 & Anvil Wisp 1 & Knot\\
\hline
U & 1.70(0.10)\footnote{Flux in units of ergs/s/cm$^2$/Hz$\times10^{-27}$.} \footnote{The first interwisp region lies between the pulsar and wisp 1. The second between wisps 1 and 2 and the third lies between wisps 2 and 3.} & 1.70(0.10)\footnote{We used $E(B-V) = 0.52$ and $R_V = 3.1$ for derreddening.} & 1.55(0.10) & 2.35(0.10) \footnote{1$\sigma$ uncertainty in the same units.} & 1.70(0.10) & 2.05(0.10) & 1.45(0.10) & -\footnote{No measurements were made.}  \\
V & 1.70(0.10) & 1.80(0.10) & 1.80(0.10) & 2.65(0.10) & 2.10(0.10) & 2.30(0.10) & 1.65(0.10) & -\\
I & 1.75(0.10) & 1.95(0.10) & 1.85(0.10) & 2.85(0.10) & 2.25(0.10) & 2.40(0.10) & 1.75(0.15) & 1.60(0.10)\\
H & 3.25(0.15) & 3.95(0.15) & 3.25(0.15) & 6.25(0.25) & 3.95(0.15)& 4.75(0.20) & 3.60(0.15) & 2.75(0.10)\\
K${\rm_s}$ & 4.05(0.15) & 4.45(0.15) & 3.70(0.15) & 6.45(0.25) & 4.05(0.15) & 4.90(0.20) & 4.10(0.15) & 2.90(0.10)\\
\hline
\hline
\end{tabular}
\end{table*}

\section{Discussion}

We used continuum filters (U$\rightarrow$K${\rm_s}$) to check the emissivity of the Crab and then compared our results with the current theoretical models.  The models from Del Zanna et al. (2003;2004;2006) seem to agree with our results. Del Zanna et al. (2006) manage to recreate the observed features seen in the Crab nebula in both X-rays and the optical, and their modelled spectrum of the PWN shows a behaviour similar to our measured spectral index of the Crab PWN.  Apart from the classical MHD approach, Del Zanna et al. (2006) added an additional equation that describes the evolution of the maximum energy of the emitting particles. This takes into account adiabatic and synchrotron losses along streamlines for the distribution of emitting particles. The spectral indices that they managed to estimate vary between $-1.15$ and $-0.6$, the variation depending on the effective magnetisation of the wind plasma.

We find that the structures within the Crab PWN (anvil and wisps) display spectra with similar spectral indices, although they are at the flatter end of the interval argued for by Del Zanna et al. Also, the regions between the wisps, despite being dimmer, also display similar red spectra. The latter suggests that the wisps exist thank to local enhancements of synchrotron emission. Two effects can be responsible for this. The first is that the wisps can be the result of an ion dominated wind that suffers a relativistic cyclotron instability (Spitkovsky \& Arons 2004) at the termination shock. The other case is that the wisps are the result of synchrotron cooling instabilities in a flow that is undergoing a transition from being particle dominated to field dominated (Hester et al. 1998;2002).

\subsection{Comparison with other pulsar wind nebulae}

There are PWNe around other pulsars that have been studied extensively. Data from Shibanov et al. (2003) for the Vela PWN, from Serafimovich et al. (2004) for the PWN of PSR B0540-69.3, and from Shibanov et al. (2008) for the PWN of 3C58 can be compared with the results for the Crab PWN. Table 6 shows the spectral indices we measured for the selected areas of the Crab PWN, along with measurements for the other PWNe.

Shibanov et al. (2003) analysed images of the Vela pulsar and its surroundings in the J$_ {\rm s}$ and H bands and detected structures around the pulsar. One structure overlaps the inner arc region seen in X-rays (Helfand et al. 2001; Pavlov et al. 2001a) and can be the IR counterpart of the X-ray emitting region. No optical confirmation has been mentioned so far, but this can be due to the redness of this region (in the same manner as in the Crab), which makes it difficult to detect. The spectral index agrees with an extension from X-rays, i.e., $\alpha_{\nu}$ in the range $-0.5$ to $-0.3$, which is marginally flatter than the spectra we derive for the inner part of the Crab PWN.

A comparison between the Crab PWN and the PWN of PSR B0540-63.9 is of particular interest since this pulsar is often referred to as the Crab twin due to its properties. Using HST images from U to I, Serafimovich et al. (2004) measured the spectral index of the PWN around PSR B0540-63.9 to be $\alpha_{\nu} \approx -1.48$ as a whole, while for various regions around the pulsar the spectral index of the PWN varies between $\alpha_{\nu} = -0.27^{+0.90}_{-0.91}$ and $-1.58^{+0.42}_{-0.43}$. The variation in the spectral index within the PWN of PSR B0540-63.9 depends on position (see Serafimovich et al. 2004 for orientation), with a tendency for the brightest areas in the optical to have the steepest spectra.

It should be kept in mind that, when comparing with the PWN of the distant PSR B0540-69.3 (for which 1\myarcsecnodot on the sky corresponds to the same true nebular size as $\sim$25\myarcsecnodot for the Crab), our Crab data probe regions much closer to the pulsar. For PSR B0540-69.3, a comparison should therefore also be made with the larger Crab survey of Veron-Cetty \& Woltjer (1993), as discussed and shown in Serafimovich et al. (2004). Nevertheless, it appears clear that the spectral index of the PWN around PSR B0540-69.3 is substantially steeper than that of the Crab PWN.

3C58 a supernova remnant similar to the Crab (e.g., plerion) has been studied from radio (Weiler \& Seiestad 1971) to IR (Shibanov et al. 2008). Using optical photometric data (B \& V) and archive data from radio to infrared, Shibanov et al. (2008) constructed a spectrum for the PWN and the torus of 3C58.  Their calculations of the spectral index for the PWN depend on the assumed extinction, but its value lies between $\alpha_\nu = -0.5$ (for the highest extinction) down to $\alpha_\nu = -1.2$. This brackets the X-ray spectral index of $\alpha_\nu = -0.88\pm0.08$, but there is no simple way to join the X-ray/optical/IR emission with a single power law. The same is true for the PSR B0540-63.9 PWN (Serafimovich et al. 2004).

 In summary, there seems to be a slight increase in steepness of the optical/NIR spectra of the PWNe of Vela, Crab and 3C58, especially if one also includes the Crab IR results of Temim et al. (2006), who find a flattening off of the Crab PWN spectrum to between $\alpha_\nu = -0.8$ and $\alpha_\nu = -0.3$ around $4\mu$m. The PWN of PSR B0540-63.9 is different from the other three in that it has a distinctly steeper spectrum, perhaps with the exception of the brightest region called ``Area 6" by Serafimovich et al. (2004). As argued by Melatos et al. (2005), the cooling time for optical/NIR photons is long compared to the flow time for the inner part of the Crab PWN. Based on the similar spectral slopes, it is likely that the same could also be true for Vela and 3C58, whereas cooling could be more efficient for most parts of the PSR B0540-63.9 PWN (see also Serafimovich et al. 2004) than for the other PWNe. 

\begin{table}
\caption{ Measued spectral indices from known PWN}
\label{expdates}
\centering
\begin{tabular}{c c c}
\hline\hline
Pulsar (Area)& Wavelength range & Spectral Index  \\
\hline
\vspace{1mm}
Crab (Wisp 1) & U$\rightarrow$ K$_ {\rm s}$ & $-0.58\pm0.08$\footnote{1$\sigma$ error in the uncertainties of the spectral indices of the Crab PWN} \\
\vspace{1mm}
Crab (Wisp 2) & U$\rightarrow$ K$_ {\rm s}$ & $-0.63\pm0.07$\\
\vspace{1mm}
Crab (Wisp 3) & U$\rightarrow$ K$_ {\rm s}$ & $-0.53\pm0.08$\\
\vspace{1mm}
Crab (Anvil Wisp) & U$\rightarrow$ K$_ {\rm s}$ & $-0.62\pm0.10$\\
\vspace{1mm}
Crab (Interwisp 1)& U$\rightarrow$ K$_ {\rm s}$ & $-0.61\pm0.08$\\
\vspace{1mm}
Crab (Interwisp 2)& U$\rightarrow$ K$_ {\rm s}$ & $-0.50\pm0.10$\\
\vspace{1mm}
Crab (Interwisp 3)& U$\rightarrow$ K$_ {\rm s}$ & $-0.49\pm0.10$\\
\vspace{1mm}
Crab (Knot)& I$\rightarrow$ K$_ {\rm s}$ & $-0.63\pm0.02$\\
\vspace{1mm}
Vela (Inner arc)\footnote{From Shibanov et al. (2003).} & X-rays$\rightarrow$IR &  $-0.5 \rightarrow -0.3$ \\
\vspace{1mm}
3C58 (Optical region)\footnote{From Shibanov et al. (2008).} & X-rays &  $-0.88\pm0.08$\\
\vspace{1mm}
3C58 & Optical$\rightarrow$IR &  $-1.2\rightarrow$-0.5\footnote{See Shibanov et al. (2008) for details.}\\
\vspace{1mm}
PSR B0540-69.3 (Area 1)\footnote{See Serafimovich et al. (2004) for orientation.} & U$\rightarrow$ I& $-1.09^{+0.30}_{-0.33}$ \\
\vspace{1mm}
PSR B0540-69.3 (Area 2) & U$\rightarrow$ I& $-1.58^{+0.42}_{-0.43}$ \\
\vspace{1mm}
PSR B0540-69.3 (Area 3) & U$\rightarrow$ I& $-1.28^{+0.41}_{-0.43}$ \\
\vspace{1mm}
PSR B0540-69.3 (Area 4) & U$\rightarrow$ I& $-1.49^{+0.59}_{-0.67}$ \\
\vspace{1mm}
PSR B0540-69.3 (Area 5) & U$\rightarrow$ I& $-0.92^{+0.33}_{-0.37}$ \\
\vspace{1mm}
PSR B0540-69.3 (Area 6) & U$\rightarrow$ I& $-0.27^{+0.90}_{-0.91}$ \\

\hline
\end{tabular}
\end{table}

\section{Conclusions}

We have presented data for the Crab pulsar in both the optical and infrared. Our infrared data suggest changes in the inner enviroment of the Crab for both dynamic structures and emissivity. Our estimated velocities for the wisps do not agree with currently known values (Hester et al. 2002). The reason for this difference can be the large time window we had. Ideally, the dynamic structure of the inner Crab should be studied using a daily based program. From our data we find that the inner knot does not show any variability in its emission and that it remains stationary over a period of 13.5 years (until 2007 December) relative to the pulsar; its average projected velocity relative to the pulsar within this time window is $\lsim 8 \kms$, corresponding to a relative space motion of $\lsim 10 \kms$, if the pulsar and knot are aligned with the jet axis. 

The combination of the optical and NIR data allowed us to create spectra of the wisps and the knot and thus obtain hints to their nature. All the wisps (and the interwisp medium) display red spectra with similar indices ($\alpha_{\nu}$ that vary between $-0.49\pm0.10$ and $-0.63\pm0.07$), and the knot has $\alpha_{\nu}=-0.63\pm0.02$. The latter seems to agree with the result of Sollerman (2003), who reported a spectral index for the knot equal to $\alpha_{\nu} = -0.8$ for a spectral range that covers both optical (U, V and I) from the HST and IR (J, H, K${\rm_s}$) from the VLT. The spectral distributions we find for the wisps and the red knot are similar to what has been estimated for other PWNe, which suggests that these observed properties are the result of the same emission mechanism. The exception appears to be the PWN of PSR B0540-63.9, which for most of its PWN has a steeper spectrum than the other PWNe and which signals stronger cooling. A feature in the Crab PWN that sticks out is the red knot with its high degree of polarisation. It lies close to the pulsar and could be part of a shock structure shared with the sprite. Both the pulsar and the sprite have markedly flatter spectra than the knot though. On a somewhat larger scale, we find that our data are compatible with the model of Del Zanna et al. (2006), although our spectra are at the flatter end of the spectral index interval argued for by Del Zanna et al. The similarity of the spectral indices of the wisps with the interwisp medium can be an indication that the wisps are just enhancements of synchrotron emission, but synchrotron cooling instabilities should not be dismissed.

We highlighted that pulsar wind nebulae of Crab like remnants display similar red spectra, and observations have also revealed that in some cases they show similar structures to those of the Crab (e.g., Vela in X-rays). Further detailed comparisons are needed to constrain the physical processes giving rise to the observed properties. Because there are many open questions regarding the nature of the PWNe, it would be interesting to extend this comparison to related objects like Crab-like remnants with bow-shock pulsars such as the Guitar Nebula, in particular to study how age affects the properties of the pulsar wind nebulae.

\begin{acknowledgements}

A. Tziamtzis would like to thank the staff of the Nordic Optical Telescope in La Palma Spain for their assistance during his stay at the NOT. The research of PL is supported by the Swedish Research Council. We thank the referee for insightful comments that improved the paper significantly.

\end{acknowledgements}

\bibliography{Crab_pulsar_astroph}

\end{document}